\documentstyle[12pt]{article}

\newcommand {\vs}[1]  { \vspace*{#1 cm} }

\newcounter{eq}
\newcounter{sc}

\newcommand {\AP}   {Ann. of Phys.}
\newcommand {\CQG}  {Class. Quantum. Grav.}

\newcommand {\IJMP}  {Int. J. Mod. Phys.}

\newcommand {\MPL}  {Mod. Phys. Lett.}
\newcommand {\NP}   {Nucl. Phys.}

\newcommand {\PL}   {Phys. Lett.}
\newcommand {\PR}   {Phys. Rev.}

\newcommand {\PTP}  {Prog. Theor. Phys.}

\def\overleftrightarrow#1{\vbox{\ialign{##\crcr
 $\leftrightarrow$\crcr\noalign{\kern-1pt\nointerlineskip}
 $\hfil\displaystyle{#1}\hfil$\crcr}}}

\newlength{\minitwocolumn}
\begin{document}

\begin{flushright}
DPUR/TH/16\\
April, 2009\\
\end{flushright}
\vspace{30pt}
\pagestyle{empty}
\baselineskip15pt

\begin{center}
{\large\bf Renormalizability of Massive Gravity in Three Dimensions
 \vskip 1mm
}

\vspace{20mm}

Ichiro Oda
          \footnote{
           E-mail address:\ ioda@phys.u-ryukyu.ac.jp
                  }

\vspace{10mm}
          Department of Physics, Faculty of Science, University of the 
           Ryukyus,\\
           Nishihara, Okinawa 903-0213, JAPAN \\

\end{center}


\vspace{20mm}
\begin{abstract}
We discuss renormalizability of a recently established, massive gravity
theory with particular higher derivative terms in three space-time dimensions. 
It is shown that this massive gravity is certainly renormalizable as well as unitary,
so it gives us a physically interesting toy model of perturbative quantum gravity 
in three dimensions.
\vspace{15mm}

\end{abstract}

\newpage
\pagestyle{plain}
\pagenumbering{arabic}


\rm
\section{Introduction}
By now it is famous that it is very difficult to construct a quantum 
theory of Einstein's general relativity in four dimensions because of 
nonrenormalizability. The nonrenormalizability is traceable to the fact
that its perturbative expansion parameter is the Einstein's gravitational
constant $\kappa$ with dimension of length squared. Thus, the divergences 
encountered in the perturbation theory of Einstein's gravity are totally
out of control and we have no physically meaningful prediction on the basis
of perturbative quantum gravity.

There is, however, an obvious renormalizable generalization of Einstein's general 
relativity in four dimensions, which is the 'higher derivative gravity' theory
with higher derivative curvature terms \cite{Stelle}, but this theory
is known to be non-unitary owing to the presence of the non-unitary massive 
$\it{ghost}$, unfortunately. 

With such a situation, it has been anticipated for a long time that
this problem might be cured by going to lower space-time dimensions,
but it is remarkable to recall that we have not yet had no grasp of 
perturbative quantum gravity, which is dynamical, unitary and possibly 
power-counting renormalizable, even in the lower dimensions.  

To date, there is only one candidate as known exceptions to this picture, which
is, what we call, topologically massive gravity in three dimensions \cite{Deser1, Deser2}. 
This geometrical gravitational theory is power-counting renormalizable despite
being dynamical, unitary and including a dimensional coupling constant like
Einstein's gravity in four dimensions. However, even this hopeful theory
has a flaw in the proof of renormalizability in the sense that there is
no gauge-invariant regularization in such a way to preserve the desirable
power-counting behavior. Related to this fact, note that dimensional
regularization cannot work well because of the explicit Levi-Civita
tensor density $\varepsilon^{\mu\nu\rho}$, gauge-invariant, higher derivative 
regulators spoil the arguments of formal renormalizability, non-covariant cutoffs
cannot be easily analyzed, and non-local regularization method involves
some assumption to be proved \cite{Deser3, Kleppe}. 

Recently, in three space-time dimensions there has been an interesting progress 
for obtaining a sensible interacting massive gravity theory \cite{Bergshoeff}
\footnote{See the references \cite{Percacci, Kaku1, Porrati, Kirsch, 
't Hooft, Kaku2, Oda1, Maeno1, Maeno2} for alternative massive gravity models.}. 
This model has been shown to be equivalent to the Pauli-Fierz massive gravity \cite{Fierz}
at the linearized approximation level and thus massive modes of helicities $\pm 2$ 
are physical propagating modes. A key idea in this model is that one adds 
higher derivative curvature terms to the Einstein-Hilbert action with the $\it{wrong}$ sign 
in such a way that the trace part of the stress-energy tensor associated with those 
higher derivative terms is proportional to the original higher derivative Lagrangian. 
With this idea, it turns out that the scalar mode coming from higher derivative Lagrangian
is precisely cancelled out \cite{Nakasone2}.
More recently, this new massive gravity model in three dimensions has been studied from
various viewpoints such as the unitarity and the impossibility of generalization to higher 
dimensions \cite{Nakasone1}, relation to the Pauli-Fierz mass term \cite{Nakasone2}, 
the AdS black hole solutions \cite{Clement}, the properties of linearized gravitational 
excitations in asymptotically AdS space-time \cite{Liu1, Liu2, Liu3} and AdS waves \cite{Eloy}. 

The aim of this article is to show that the new massive gravity theory in three dimensions
\cite{Bergshoeff} is indeed renormalizable within the framework of perturbation theory. 
It is worhtwhile to notice here that we can make use of dimensional regularization 
in the new massive gravity theory since there is no the Levi-Civita tensor density 
$\varepsilon^{\mu\nu\rho}$ in the action, which should be constrasted to the case of the 
topologically massive gravity. 

The reason why the new massive gravity is renormalizable is very simple. With a suitable choice 
of the regulator and gauge condition for diffeomorphisms, the graviton propagator falls off 
like $\frac{1}{p^4}$, implying that the gravitational field has the ultraviolet dimension 
of $(mass)^{-\frac{1}{2}}$. Then, the power counting argument together with the Slavnov-Taylor 
identity and renormalization equation for the effective action shows that the divergent part 
of the effective action consists of terms of dimension $3$ at most and is invariant under 
diffeomorphisms. Thus, only the admissible counter-terms are the cosmological term and 
the Einstein-Hilbert action, and thereby the theory is perturbatively renormalizable. 
Note that the higher derivative curvature-squared terms receive no corrections so that 
the property of the unitarity in the new massive gravity is not violated even after 
the renormalization procedure. 

Since it has been already shown that this massive gravity is unitary, we have for the first time 
a physically plausible quantum theory of gravity within the framework of perturbation theory
such that it is dynamical, renormalizable and unitary although this theory can be formulated 
only in three dimensions. 

In the next section, we perform the BRST quantization of the new massive gravity theory 
in three dimensions. 
In the third section, we derive the propagator for the gravitational field on the
basis of the gauge fixed, BRST-invariant action obtained in the section 2.
In the fourth section, we consider the power-counting and determine the
divergent part of the effective action by using the Slavnov-Taylor identity
and the renormalization equation.
The final section is devoted to conclusion and discussions.

\section{BRST quantization of the new massive gravity theory}

We start with BRST quantization of the new massive gravity theory
in three dimensions \cite{Bergshoeff} which is an interactive and unitary theory 
with higher derivative terms for the massive graviton. The classical action
which we consider is of form \footnote{The space-time indices $\mu, \nu, 
\cdots$ run over $0, 1, 2$. We take the metric signature $(-, +, +)$ and 
follow the notation and conventions of the textbook of MTW \cite{MTW}.}:  
\begin{eqnarray}
S_c &=& \int d^3 x {\cal{L}}_c
\nonumber\\
&=& \int d^3 x \frac{1}{\kappa^2} \sqrt{- g} [ - R  
+ \frac{1}{M^2} ( R_{\mu\nu} R^{\mu\nu} - \frac{3}{8} R^2 ) ],
\label{Action1}
\end{eqnarray}
where $\kappa^2 \equiv 16 \pi G$ ($G$ is the $3$-dimensional Newton's constant) and
$M$ is a constant of mass dimension.
Let us note that $\kappa$ has dimension of $(mass)^{-\frac{1}{2}}$, so
the theory defined by the action (\ref{Action1}) might at first sight appear to be
unrenormalizable. As usual let us write
\begin{eqnarray}
g_{\mu\nu} = \eta_{\mu\nu} + \kappa h_{\mu\nu},
\label{G-field}
\end{eqnarray}
where a flat Minkowski background $\eta_{\mu\nu}$ has the diagonal 
element $(-1, 1, 1)$. With this definition, the gravitational field
$h_{\mu\nu}$ has canonical dimension of $(mass)^{\frac{1}{2}}$.
Consequently, the Einstein-Hilbert action and the graviton-matter interaction terms
have canonical dimensions greater than three, which is the origin of
unrenormalizability of Einstein's general relativity without the higher derivative
terms. Later, we will see that the gravitational field $h_{\mu\nu}$ has ultraviolet 
dimension of $(mass)^{- \frac{1}{2}}$, whose fact leads to renormalizability 
of the new massive gravity theory under consideration.

Now the BRST transformations for the metric tensor, ghost, antighost and
Nakanishi-Lautrup auxiliary field are respectively given by
\begin{eqnarray}
\delta_B g_{\mu\nu} &=& - \kappa^3 ( \nabla_\mu c_\nu + \nabla_\nu c_\mu ), 
\nonumber\\
\delta_B c^\mu &=& - \kappa^3 c^\nu \partial_\nu c^\mu, 
\nonumber\\
\delta_B \bar c_\mu &=& i b_\mu,
\nonumber\\
\delta_B b_\mu &=& 0,
\label{BRST}
\end{eqnarray}
where the covariant derivative is defined as usual by
$\nabla_\mu c_\nu = \partial_\mu c_\nu - \Gamma^\lambda_{\mu\nu} c_\lambda$
with the affine connection $\Gamma^\lambda_{\mu\nu}$. It is straightforward to 
prove that the BRST transformations (\ref{BRST}) are off-shell nilpotent.

Next, we wish to fix the gauge invariance by some suitable gauge fixing
condition. The usual gauge fixing condition would be the de Donder's gauge
(or its higher derivative generalizations) because it is known that
the renormalization procedure becomes extremely simple in this gauge as first pointed out 
by Stelle \cite{Stelle}.  However, it turns out that only the de Donder's gauge
does not work well in the present theory owing to the scalar mode whose propagator 
$\it{does}$ behave as $\frac{1}{p^2}$ unlike $\frac{1}{p^4}$ for large momenta.

It is clear that this scalar mode is nothing but the $\it{conformal}$ mode stemming from 
the Einstein-Hilbert action since the conformal mode from the higher derivative
curvature terms is exactly cancelled out by a specific combination of them. 
The similar situation has been already happened in the case of the topologically massive gravity
theory where there exists the conformal mode only from the Einstein-Hilbert action
since the gravitational Chern-Simons term, which is of higher-derivative nature, is conformally 
invariant so that there is no conformal mode in this term \cite{Deser3}. 
Indeed, the existence of the scalar mode has also caused some trouble in the proof of renormalizability
in the topologically massive gravity theory.
 However, at the linearized level the new massive gravity is shown to be equivalent to the 
Pauli-Fierz massive gravity with only the spin $\pm 2$ massive graviton modes without the scalar mode 
by path integral \footnote{But there has not yet been the proof of this statement by canonical formalism.},
so the troublesome scalar mode might be somewhat an artificial degree of freedom.

In order to resolve this problem associated with the propagator for the scalar mode, 
we shall obey the following line of the argument. First, the key observation is to recall 
that provided that we add the Pauli-Fierz mass term 
\begin{eqnarray}
{\cal{L}}_{PF} = - \frac{m^2}{4} ( h_{\mu\nu} h^{\mu\nu} - h^2 ),
\label{PF}
\end{eqnarray}
to the action (\ref{Action1}), we have no propagator for the scalar mode while the propagator 
for the massive gravitational modes of helicities $\pm 2$ falls off like $\frac{1}{p^4}$ 
for large momenta \cite{Nakasone2} as desired. However, if we simply added the Pauli-Fierz mass term 
to the action (\ref{Action1}) without care, the BRST invariance would be broken by the additional 
term and as a result the unitarity would be violated whose situation we have to avoid from physical fundamental
principle. Moreover, it is obvious that addition of the the Pauli-Fierz mass term would modify
the physical content of the original theory.

Accordingly, we shall add the Pauli-Fierz mass term to the action (\ref{Action1}) as some (infrared)
regulator (which means that we take the vanishing mass limit after renormalization) in a BRST-invariant
manner. Such a BRST procedure has been already studied from a different interest \cite{Hamamoto}, 
but the method turns out to be applicable to the present problem as well.
Using the BRST procedure in Ref. \cite{Hamamoto}, it is easy to obtain the following gauge fixed,
BRST-invariant action
\begin{eqnarray}
S &\equiv& \int d^3 x {\cal{L}}
\nonumber\\
&=& \int d^3 x [ \frac{1}{\kappa^2} \sqrt{- g} \{ - R  
+ \frac{1}{M^2} ( R_{\mu\nu} R^{\mu\nu} - \frac{3}{8} R^2 ) \}
+ \frac{1}{\kappa^3} \tilde g^{\mu\nu} \partial_\mu b_\nu
\nonumber\\
&+& i \partial_\mu \bar c_\nu D_\rho^{\mu\nu} c^\rho - \frac{m^2}{4} ( h_{\mu\nu} h^{\mu\nu} - h^2 )
- \frac{1}{4} F_{\mu\nu}^2 - \frac{1}{\kappa^2} b \partial_\mu A^\mu
\nonumber\\
&-& ( \partial^\nu h_{\mu\nu} - \partial_\mu h ) ( m A^\mu - \partial^\mu \varphi )
- i \bar c ( \Box c - m \partial_\mu c^\mu ) ],
\label{BRST-Action}
\end{eqnarray}
where we have defined $\tilde g^{\mu\nu} = \sqrt{-g} g^{\mu\nu}$, 
$D_\rho^{\mu\nu} = \tilde g^{\mu\sigma} \delta_\rho^\nu \partial_\sigma
+ \tilde g^{\nu\sigma} \delta_\rho^\mu \partial_\sigma 
- \tilde g^{\mu\nu} \partial_\rho - (\partial_\rho \tilde g^{\mu\nu})$
and $F_{\mu\nu} = \partial_\mu A_\nu - \partial_\nu A_\mu$.
Furthermore, in order to get the BRST symmetric theory, the Steuckelberg 
fields $A^\mu$ and $\varphi$, and the new ghost $c$, the corresponding antighost 
$\bar c$ and Nakanishi-Lautrup field $b$ are introduced. In addition to
Eq. (\ref{BRST}), the BRST transformations for them are given by
\begin{eqnarray}
\delta_B A^\mu &=& \kappa^2 ( - m c^\mu + \partial^\mu c ), 
\nonumber\\
\delta_B \varphi &=& \kappa^2 m  c, 
\nonumber\\
\delta_B \bar c &=& i b,
\nonumber\\
\delta_B b &=& \delta_B c = 0,
\label{BRST2}
\end{eqnarray}
where note in particular that we can set $\delta_B^2 A^\mu = 0$ since 
$\delta_B^2 A^\mu = \kappa^5 m c^\nu \partial_\nu c^\mu$ belongs to the
higher-order in $\kappa$.

At this stage, we should comment on the meaning of the above action (\ref{BRST-Action}) in detail.
As the gauge fixing conditions of diffeomorphisms and Steuckelberg-like gauge symmetry,
we have selected the de Donder's gauge $ \partial_\mu \tilde g^{\mu\nu} = 0$
and the Landau-like gauge $\partial_\mu A^\mu = 0$. Then, the terms containing the fields
$A_\mu, \varphi, b, c$ and $\bar c$ are only quadratic in fields and therefore there are no
interaction terms among them. In obtaining the propagators, off-diagonal pieces such as
$< b_\mu h^{\nu\rho} >$, $< b A^\mu >$ and $< h_{\mu\nu} A^\rho >$ are therefore irrelevant
as $b_\mu$, $b$ and $A^\mu$ never appear in vertices.

As seen shortly in the next section, it will turn out that the propagator
for the scalar (conformal) mode does not exist in the action (\ref{BRST-Action}). Then,
one asks ourselves where one dynamical degree of freedom associated with the scalar
mode has gone away since the number of dynamical degrees of freedom never changes in perturbation 
theory. The answer lies in the fact that there appears the propagator for the 
Steuckelberg field $A_\mu$, which has one dynamical degree of freedom owing to gauge
invariance. In other words, because of the BRST-invariant regulator the massless pole 
$\frac{1}{p^2}$ of the scalar mode is transferred to that of the Steuckelberg field $A_\mu$. 
This fact could be understood more clearly by taking the more general gauge condition 
$\partial_\mu A^\mu - \frac{m}{2} h = 0$. 

Finally, in the limit of $m \rightarrow 0$, the action (\ref{BRST-Action})
reduces to the gauge fixed and BRST-invariant theory of the original new massive gravity 
up to irrelevant non-interacting terms. This limit must be taken after the whole renormalization 
procedure is completed.

\section{Graviton propagator}

On the basis of the gauge fixed, BRST-invariant action (\ref{BRST-Action}),
we wish to derive the propagator for the gravitational field $h_{\mu\nu}$.
To this end, it is useful to take account of the spin projection operators 
in $3$ space-time dimensions \cite{Stelle}. A set of the spin operators 
$P^{(2)}, P^{(1)}, P^{(0, s)}, P^{(0, w)}, P^{(0, sw)}$
and  $P^{(0, ws)}$ form a complete set in the space of second rank symmetric 
tensors and are defined as
\begin{eqnarray}
P^{(2)}_{\mu\nu, \rho\sigma} &=& \frac{1}{2} ( \theta_{\mu\rho} \theta_{\nu\sigma}
+ \theta_{\mu\sigma} \theta_{\nu\rho} ) - \frac{1}{2} \theta_{\mu\nu} \theta_{\rho\sigma},
\nonumber\\  
P^{(1)}_{\mu\nu, \rho\sigma} &=& \frac{1}{2} ( \theta_{\mu\rho} \omega_{\nu\sigma}
+ \theta_{\mu\sigma} \omega_{\nu\rho} + \theta_{\nu\rho} \omega_{\mu\sigma}
+ \theta_{\nu\sigma} \omega_{\mu\rho} ),
\nonumber\\  
P^{(0, s)}_{\mu\nu, \rho\sigma} &=& \frac{1}{2} \theta_{\mu\nu} \theta_{\rho\sigma},
\nonumber\\  
P^{(0, w)}_{\mu\nu, \rho\sigma} &=& \omega_{\mu\nu} \omega_{\rho\sigma},
\nonumber\\  
P^{(0, sw)}_{\mu\nu, \rho\sigma} &=& \frac{1}{\sqrt{2}} \theta_{\mu\nu} \omega_{\rho\sigma},
\nonumber\\  
P^{(0, ws)}_{\mu\nu, \rho\sigma} &=& \frac{1}{\sqrt{2}} \omega_{\mu\nu} \theta_{\rho\sigma}.
\label{Spin projectors}
\end{eqnarray}
Here the transverse operator $\theta_{\mu\nu}$ and the longitudinal operator
$\omega_{\mu\nu}$ are defined as
\begin{eqnarray}
\theta_{\mu\nu} &=& \eta_{\mu\nu} - \frac{1}{\Box} \partial_\mu \partial_\nu 
= \eta_{\mu\nu} - \omega_{\mu\nu}, \nonumber\\  
\omega_{\mu\nu} &=& \frac{1}{\Box} \partial_\mu \partial_\nu.
\label{theta}
\end{eqnarray}
It is straightforward to show that the spin projection operators satisfy 
the orthogonality relations
\begin{eqnarray}
P_{\mu\nu, \rho\sigma}^{(i, a)} P_{\rho\sigma, \lambda\tau}^{(j, b)}
&=& \delta^{ij} \delta^{ab} P_{\mu\nu, \lambda\tau}^{(i, a)},
\nonumber\\  
P_{\mu\nu, \rho\sigma}^{(i, ab)} P_{\rho\sigma, \lambda\tau}^{(j, cd)}
&=& \delta^{ij} \delta^{bc} P_{\mu\nu, \lambda\tau}^{(i, a)},
\nonumber\\  
P_{\mu\nu, \rho\sigma}^{(i, a)} P_{\rho\sigma, \lambda\tau}^{(j, bc)}
&=& \delta^{ij} \delta^{ab} P_{\mu\nu, \lambda\tau}^{(i, ac)},
\nonumber\\  
P_{\mu\nu, \rho\sigma}^{(i, ab)} P_{\rho\sigma, \lambda\tau}^{(j, c)}
&=& \delta^{ij} \delta^{bc} P_{\mu\nu, \lambda\tau}^{(i, ac)},
\label{Orthogonality}
\end{eqnarray}
with $i, j = 0, 1, 2$ and $a, b, c, d = s, w$ and the tensorial relation 
\begin{eqnarray}
[ P^{(2)} + P^{(1)} + P^{(0, s)} + P^{(0, w)} ]_{\mu\nu, \rho\sigma} = 
\frac{1}{2} ( \eta_{\mu\rho}\eta_{\nu\sigma} + \eta_{\mu\sigma}\eta_{\nu\rho} ).
\label{T relation}
\end{eqnarray}

It is then straightforward to extract the quadratic fluctuations in 
$h_{\mu\nu}$ from each term of the action (\ref{BRST-Action}) and express it 
in terms of the spin projection operators:
\begin{eqnarray}
{\cal{L}}_{EH} &\equiv& \sqrt{- g} R 
\nonumber\\
&=& \frac{\kappa^2}{4} h^{\mu\nu} [ P^{(2)} - P^{(0, s)} ]_{\mu\nu, \rho\sigma}
\Box h^{\rho\sigma},
\nonumber\\
{\cal{L}}_{R^2} &\equiv& \sqrt{- g} R^2 
\nonumber\\
&=& 2 \kappa^2 h^{\mu\nu} P^{(0, s)}_{\mu\nu, \rho\sigma}
\Box^2 h^{\rho\sigma},
\nonumber\\
{\cal{L}}_{R_{\mu\nu}^2} &\equiv& \sqrt{- g} R_{\mu\nu} R^{\mu\nu} 
\nonumber\\
&=&  \frac{\kappa^2}{4} h^{\mu\nu} [ P^{(2)} + 3 P^{(0, s)} ]_{\mu\nu, \rho\sigma}
\Box^2 h^{\rho\sigma},
\nonumber\\
{\cal{L}}_{PF} &\equiv& - \frac{m^2}{4} ( h_{\mu\nu} h^{\mu\nu} - h^2 )
\nonumber\\
&=& - \frac{m^2}{4} h^{\mu\nu} [ P^{(2)} + P^{(1)} - P^{(0, s)} 
- \sqrt{2} ( P^{(0, sw)} + P^{(0, ws)} ) ]_{\mu\nu, \rho\sigma} h^{\rho\sigma}.
\label{Lagrangian}
\end{eqnarray}
Note that a salient feature of the specific combination of the higher derivative terms
is the disappearance of the spin projection operator corresponding to the spin 0 scalar mode:
\begin{eqnarray}
\sqrt{- g} ( R_{\mu\nu} R^{\mu\nu} - \frac{3}{8} R^2 )
= \frac{\kappa^2}{4} h^{\mu\nu} P^{(2)}_{\mu\nu, \rho\sigma}
\Box^2 h^{\rho\sigma}.
\label{Peculiar relation}
\end{eqnarray}

Using the relations (\ref{Lagrangian}) and (\ref{Peculiar relation}), 
the quadratic part in $h_{\mu\nu}$ of the action (\ref{BRST-Action}) is expressed
in term of the spin projection operators like
\begin{eqnarray}
S = \int d^3 x \frac{1}{2} h^{\mu\nu} {\cal{P}}_{\mu\nu, \rho\sigma} 
h^{\rho\sigma},
\label{Action5}
\end{eqnarray}
where ${\cal{P}}_{\mu\nu, \rho\sigma}$ is defined as
\begin{eqnarray}
{\cal{P}}_{\mu\nu, \rho\sigma} &=& [ \frac{1}{2} ( \frac{1}{M^2} \Box^2
- \Box - m^2 ) P^{(2)} - \frac{m^2}{2} P^{(1)} 
+ \frac{1}{2} ( \Box + m^2 ) P^{(0, s)}
\nonumber\\
&+& \frac{m^2}{\sqrt{2}} ( P^{(0, sw)} + P^{(0, ws)} )]_{\mu\nu, \rho\sigma}.
\label{P}
\end{eqnarray}
Then, the propagator for $h_{\mu\nu}$ is defined by
\begin{eqnarray}
<0| T (h_{\mu\nu}(x) h_{\rho\sigma}(y)) |0>
= i {\cal{P}}_{\mu\nu, \rho\sigma}^{-1} \delta^{(3)}(x-y),
\label{Propa}
\end{eqnarray}
where using the relations (\ref{Orthogonality}) and (\ref{T relation}), 
the inverse of the operator $\cal{P}$ is easily calculated as
\begin{eqnarray}
{\cal{P}}_{\mu\nu, \rho\sigma}^{-1} 
&=& [ \frac{2}{\frac{1}{M^2} \Box^2 - \Box - m^2} P^{(2)} 
-  \frac{2}{m^2} P^{(1)} 
- \frac{\Box + m^2}{m^4} P^{(0, w)} 
\nonumber\\
&+& \frac{\sqrt{2}}{m^2} ( P^{(0, sw)} + P^{(0, ws)} )]_{\mu\nu, \rho\sigma}.
\label{P-inv}
\end{eqnarray}

This expression of the graviton propagator brings us a few important information.
First, let us recall that the spin 1 component projected by $P^{(1)}$ and 
the spin 0 ones by $P^{(0, w)}, P^{(0, sw)}$, and  $P^{(0, ws)}$ can be gauged away,
whereas the spin 2 component projected by $P^{(2)}$ and the spin 0 one
projected by $P^{(0, s)}$ are physically relevant. One notable feature
of the propagator (\ref{P-inv}) is that there is no spin 0 component
projected by $P^{(0, s)}$, which is the reason why we have taken the 
Pauli-Fierz mass term as the regulator of the conformal mode. Second, the
propagator of the spin 2 massive graviton modes of helicities $\pm 2$
falls off like $\frac{1}{p^4}$ for large momenta. 

Finally, let us notice that as the result of addition of the Pauli-Fierz mass
term, there appear two massive poles in the sector of spin 2 
graviton, which is of form  
\begin{eqnarray}
I &\equiv& \frac{1}{M^2} \Box^2 - \Box - m^2
\nonumber\\
&=& \frac{1}{M^2} ( \Box - \omega_+ ) ( \Box - \omega_- ),
\label{Pole1}
\end{eqnarray}
where $\omega_{\pm} \equiv \frac{1 \pm \sqrt{1 + 4 (\frac{m}{M})^2}}
{2} M^2$, which are real numbers such that $\omega_+ > 0, \omega_- < 0$ \cite{Nakasone2}.
As examined in Ref. \cite{Nakasone2}, the presence of the negative pole $\omega_-$
might seem to induce violation of both unitarity and causality. But such a
pathology now does not occur since in the $m \rightarrow 0$ limit, the positive one of 
the two massive poles, $\omega_+$ remains a pole of massive graviton while the other 
negative pole $\omega_-$ changes to a harmless pole of massless graviton.
In this respect, it is worth mentioning that the original massive gravity theory has 
been already shown to be unitary and causal in the de Donder's gauge \cite{Nakasone1}.

\section{Structure of the divergences}

We now turn our attention to the analysis of structure of the divergences
of the effective action.
To this aim, let us first add sources $K_{\mu\nu}$ (anti-commuting, ghost
number = $-1$, dimension = $\frac{3}{2}$), $L_\mu$ (commuting, ghost number = $-2$, 
dimension = $1$), $M_\mu$ (anti-commuting, ghost number = $-1$, dimension = $1$)
and $N$ (anti-commuting, ghost number = $-1$, dimension = $1$) for the BRST 
transformations of $\tilde g^{\mu\nu}$, $c^\mu$, $A^\mu$ and $\varphi$ to the 
action (\ref{BRST-Action}), respectively \footnote{Here for simplicity we have
regarded $\tilde g^{\mu\nu}$ as a basic gravitational field.}:
\begin{eqnarray}
\tilde S &\equiv& \int d^3 x \tilde \Sigma 
\nonumber\\
&=& \int d^3 x [ {\cal{L}}
+ \frac{1}{\kappa^3} K_{\mu\nu} \delta_B \tilde g^{\mu\nu} 
+ \frac{1}{\kappa^3} L_\mu \delta_B c^\mu 
+ \frac{1}{\kappa^3} M_\mu \delta_B A^\mu 
+ \frac{1}{\kappa^3} N \delta_B \varphi ]
\nonumber\\
&=& \int d^3 x [ \frac{1}{\kappa^2} \sqrt{- g} \{ - R  
+ \frac{1}{M^2} ( R_{\mu\nu} R^{\mu\nu} - \frac{3}{8} R^2 ) \}
+ \frac{1}{\kappa^3} \tilde g^{\mu\nu} \partial_\mu b_\nu
+ ( K_{\mu\nu} + i \partial_\mu \bar c_\nu ) D_\rho^{\mu\nu} c^\rho
\nonumber\\
&-& \frac{m^2}{4} ( h_{\mu\nu} h^{\mu\nu} - h^2 ) - \frac{1}{4} F_{\mu\nu}^2
- \frac{1}{\kappa^2} b \partial_\mu A^\mu
- ( \partial^\nu h_{\mu\nu} - \partial_\mu h ) ( m A^\mu - \partial^\mu \varphi )
\nonumber\\
&-& i \bar c ( \Box c - m \partial_\mu c^\mu ) 
- L_\mu c^\nu \partial_\nu c^\mu 
+ \frac{1}{\kappa} M_\mu ( -m c^\mu + \partial^\mu c ) + \frac{m}{\kappa} N c ].
\label{BRST-Action with sources}
\end{eqnarray}

Next, based on this action, let us consider the superficial degree of divergence
for 1PI (one particle irreducible) Feynman diagrams. 
Then, it is convenient to introduce the following notation:
$n_R =$ number of graviton vertices with two derivatives, $n_{R^2} =$ number 
of graviton vertices with four derivatives, $n_G =$ number of ghost vertices,
$n_K =$ number of $K$-graviton-ghost vertices, $n_L =$ number of $L$-ghost-ghost vertices,
$I_G =$ number of internal ghost propagators and $I_E =$ number of internal
graviton propagators. Note that the fields $A_\mu, \varphi, b, c$ and $\bar c$ are 
free so we can exclude such the fields from the counting of the superficial degree of 
divergence. Using this fact and the above notation, the superficial degree of divergence 
for an arbitrary Feynman diagram $\gamma$ can be easily calculated to be
\begin{eqnarray}
\omega(\gamma) &\equiv& \sum n_i d_i + (3-4) I_E + (3-2) I_G
- 3 (\sum n_i - 1)
\nonumber\\
&=& 3 - n_R + n_{R^2} - n_G - 2 n_K - 2 n_L - I_E + I_G,
\label{SDD1}
\end{eqnarray}
where $d_i$ denotes the number of derivatives in the interaction terms.
Here we have made use of the fact that graviton propagator behaves like $p^{-4}$
for large momenta as mentioned in the previous section.

Furthermore, using the relation
\begin{eqnarray}
2 n_G + 2 n_L + n_K = 2 I_G + E_c + E_{\bar c},
\label{Ghost relation}
\end{eqnarray}
with the notation that $E_c =$ number of external ghosts $c^\mu$ and
$E_{\bar c} =$ number of external antighosts $\bar c_\mu$,
Eq. (\ref{SDD1}) is cast to the form
\begin{eqnarray}
\omega(\gamma) = 3 - n_R - ( I_E - n_{R^2} ) - \frac{3}{2} n_K -  n_L 
- \frac{1}{2} E_c -  \frac{1}{2} E_{\bar c}.
\label{SDD2}
\end{eqnarray}

However, as noticed before by Stelle \cite{Stelle}, the graviton propagator
vanishes when multiplied by the momenta $p^\mu$, so the terms with the 
de Donder's gauge term $\partial_\mu \tilde g^{\mu\nu}$ do not connect with
the graviton propagator, and thus we can neglect such the terms. 
Consequently, since the dimension of the ghost and antighost is increased by
2, instead of (\ref{SDD2}), the correct superficial degree of divergence
is given by
\begin{eqnarray}
\omega(\gamma) = 3 - n_R - ( I_E - n_{R^2} ) - \frac{3}{2} n_K -  n_L 
- \frac{5}{2} E_c -  \frac{5}{2} E_{\bar c}.
\label{SDD3}
\end{eqnarray}

The superficial degree of divergence (\ref{SDD3}) gives rise to
some useful information. First, since there is a relation $I_E \geq n_{R^2}$ 
for 1PI diagrams $\gamma$, we have
\begin{eqnarray}
\omega(\gamma) \le 3,
\label{SDD4}
\end{eqnarray}
which indicates that the bound on the superficial degree of divergence
for 1PI diagrams $\gamma$ is cubic to all orders, and the theory is power-counting
renormalizable. In this context, it is worthwhile to point out that we can
make use of the dimensional regularization in evaluating various
amplitudes if necessary. 

Second, we can understand a convergent property of some 1PI Feynman
digrams via Eq. (\ref{SDD3}). In order to describe this fact in the physical terminology, 
it is useful to define the effective action $\Gamma$ of this theory in a conventional
way and expand it in a power series of the loop number like
\begin{eqnarray}
\Gamma = \Gamma^{(0)} + \Gamma^{(1)} + \Gamma^{(2)} + \cdots, 
\label{EA}
\end{eqnarray}
where $\Gamma^{(n)}$ describes the $n$-loop part of the effective action.
Moreover, we separate $\Gamma^{(n)}$ into the divergent and finite parts like
\begin{eqnarray}
\Gamma^{(n)} = \Gamma_{div}^{(n)} + \Gamma_{fin}^{(n)}.
\label{Separation}
\end{eqnarray}
Now notice that all the 1PI diagrams involving external ghosts or K-vertices or 
L-ones are finite since, for instance, if $n_K = 1$, we have $E_c = 1$, and thus 
Eq. (\ref{SDD3}) leads to $\omega(\gamma) \le -1$.
Hence, we can describe these convergent properties in terms of the divergent part 
of $\Gamma^{(n)}$ as
\begin{eqnarray}
\frac{\delta \Gamma_{div}^{(n)}}{\delta c^\sigma}
= \frac{\delta \Gamma_{div}^{(n)}}{\delta K_{\mu\nu}} 
= \frac{\delta \Gamma_{div}^{(n)}}{\delta L_\mu} 
= 0. 
\label{Div-Gamma}
\end{eqnarray}

Next we move on to the Slavnov-Taylor identity. The BRST invariance of the action 
(\ref{BRST-Action with sources}) enables us to derive the Slavnov-Taylor identity for 
$\tilde \Sigma$ as follows:
\begin{eqnarray}
0 &=& \delta_B \tilde \Sigma
\nonumber\\
&=& \kappa^3 D_\rho^{\mu\nu} c^\rho \frac{\delta \tilde \Sigma}{\delta \tilde g^{\mu\nu}} 
- \kappa^3 c^\nu \partial_\nu c^\sigma \frac{\delta \tilde \Sigma}{\delta c^\sigma}
+ i b_\tau \frac{\delta \tilde \Sigma}{\delta \bar c_\tau}
\nonumber\\
&+& \kappa^2 ( -m c^\mu + \partial^\mu c ) \frac{\delta \tilde \Sigma}{\delta A^\mu}
+ \kappa^2 m c \frac{\delta \tilde \Sigma}{\delta \varphi}
+ i b \frac{\delta \tilde \Sigma}{\delta \bar c}.
\label{ST1}
\end{eqnarray}
With the help of relations
\begin{eqnarray}
\frac{\delta \tilde \Sigma}{\delta K_{\mu\nu}} &=&  D_\rho^{\mu\nu} c^\rho, \
\frac{\delta \tilde \Sigma}{\delta L_\sigma} = - c^\nu \partial_\nu c^\sigma,
\nonumber\\ 
\frac{\delta \tilde \Sigma}{\delta M_\mu} &=& \frac{1}{\kappa} ( -m c^\mu + \partial^\mu c ), 
\
\frac{\delta \tilde \Sigma}{\delta N} = \frac{m}{\kappa} c,
\label{Relations}
\end{eqnarray}
Eq. (\ref{ST1}) can be rewritten as
\begin{eqnarray}
0 &=& \delta_B \tilde \Sigma
\nonumber\\
&=& \kappa^3  \frac{\delta \tilde \Sigma}{\delta K_{\mu\nu}} 
\frac{\delta \tilde \Sigma}{\delta \tilde g^{\mu\nu}} 
+ \kappa^3 \frac{\delta \tilde \Sigma}{\delta L_\sigma}
\frac{\delta \tilde \Sigma}{\delta c^\sigma}
+ i b_\tau \frac{\delta \tilde \Sigma}{\delta \bar c_\tau}
\nonumber\\
&+& \kappa^3 \frac{\delta \tilde \Sigma}{\delta M_\mu} 
\frac{\delta \tilde \Sigma}{\delta A^\mu}
+ \kappa^3 \frac{\delta \tilde \Sigma}{\delta N} 
\frac{\delta \tilde \Sigma}{\delta \varphi}
+ i b \frac{\delta \tilde \Sigma}{\delta \bar c}.
\label{ST2}
\end{eqnarray}

If we define $\Sigma$ by
\begin{eqnarray}
\Sigma = \tilde \Sigma - \int d^3 x 
[ \frac{1}{\kappa^3} \tilde g^{\mu\nu} \partial_\mu b_\nu
- \frac{1}{\kappa^2} b \partial_\mu A^\mu ],
\label{Sigma}
\end{eqnarray}
we can obtain the Slavnov-Taylor identity for $\Sigma$:
\begin{eqnarray}
\frac{\delta \Sigma}{\delta K_{\mu\nu}} 
\frac{\delta \Sigma}{\delta \tilde g^{\mu\nu}} 
+ \frac{\delta \Sigma}{\delta L_\sigma}
\frac{\delta \Sigma}{\delta c^\sigma}
+ \frac{\delta \Sigma}{\delta M_\mu} 
\frac{\delta \Sigma}{\delta A^\mu}
+ \frac{\delta \Sigma}{\delta N} 
\frac{\delta \Sigma}{\delta \varphi}
= 0,
\label{ST3}
\end{eqnarray}
where we have made use of the equations of motion to the antighosts:
\begin{eqnarray}
i \partial_\mu \frac{\delta \Sigma}{\delta K_{\mu\nu}}
+  \frac{\delta \Sigma}{\delta \bar c_\nu}
&=& 0,
\nonumber\\
i \kappa \partial_\mu \frac{\delta \Sigma}{\delta M_\mu}
+  \frac{\delta \Sigma}{\delta \bar c}
&=& 0.
\label{Antighost eq.}
\end{eqnarray}

Because at the zero-loop order we have a relation $\Gamma^{(0)} = \Sigma$,
it is expected that the ST-identity for the effective action takes the same form 
as (\ref{ST3}) where $\Sigma$ is now replaced with $\Gamma$. In fact, this statement 
can be easily verified. Thus we arrive at the the ST-identity for the effective action
\begin{eqnarray}
\frac{\delta \Gamma}{\delta K_{\mu\nu}} 
\frac{\delta \Gamma}{\delta \tilde g^{\mu\nu}} 
+ \frac{\delta \Gamma}{\delta L_\sigma}
\frac{\delta \Gamma}{\delta c^\sigma}
+ \frac{\delta \Gamma}{\delta M_\mu} 
\frac{\delta \Gamma}{\delta A^\mu}
+ \frac{\delta \Gamma}{\delta N} 
\frac{\delta \Gamma}{\delta \varphi}
= 0.
\label{ST-identity}
\end{eqnarray}

Let us recall that the ST-identity for the effective action naturally leads to the 
renormalization equation to the the divergent part $\Gamma^{(n)}$ of the effective action
\begin{eqnarray}
{\cal{G}}  \Gamma_{div}^{(n)} = 0,
\label{RE1}
\end{eqnarray}
where the operator ${\cal{G}}$ is defined as
\begin{eqnarray}
{\cal{G}} 
&=& \frac{\delta \Gamma^{(0)}}{\delta K_{\mu\nu}} \frac{\delta}{\delta \tilde g^{\mu\nu}}
+ \frac{\delta \Gamma^{(0)}}{\delta L_\sigma} \frac{\delta}{\delta c^\sigma}
+ \frac{\delta \Gamma^{(0)}}{\delta M_\mu} \frac{\delta}{\delta A^\mu}
+ \frac{\delta \Gamma^{(0)}}{\delta N} \frac{\delta}{\delta \varphi}
\nonumber\\
&+& \frac{\delta \Gamma^{(0)}}{\delta \tilde g^{\mu\nu}} \frac{\delta}{\delta K_{\mu\nu}}
+ \frac{\delta \Gamma^{(0)}}{\delta c^\sigma} \frac{\delta}{\delta L_\sigma}
+ \frac{\delta \Gamma^{(0)}}{\delta A^\mu} \frac{\delta}{\delta M_\mu}
+ \frac{\delta \Gamma^{(0)}}{\delta \varphi} \frac{\delta}{\delta N}.
\label{G-operator}
\end{eqnarray}

Using the relations (\ref{Div-Gamma}) and the obvious identities
\begin{eqnarray}
\frac{\delta \Gamma_{div}^{(n)}}{\delta A^\mu} 
= \frac{\delta \Gamma_{div}^{(n)}}{\delta \varphi} 
= \frac{\delta \Gamma_{div}^{(n)}}{\delta M_\mu} 
= \frac{\delta \Gamma_{div}^{(n)}}{\delta N} 
= 0,
\label{Identities}
\end{eqnarray}
the renormalization equation becomes
\begin{eqnarray}
0 =  {\cal{G}}  \Gamma_{div}^{(n)}
= \frac{\delta \Gamma^{(0)}}{\delta K_{\mu\nu}} 
\frac{\delta \Gamma_{div}^{(n)}}{\delta \tilde g^{\mu\nu}}
= D_\rho^{\mu\nu} c^\rho \frac{\delta \Gamma_{div}^{(n)}}{\delta \tilde g^{\mu\nu}}. 
\label{RE2}
\end{eqnarray}
This renormalization equation clearly shows that $\Gamma_{div}^{(n)}$ is
a BRST-invariant functional of $\tilde g^{\mu\nu}$. Since the divergent part is
local and of dimension 3 at most, the only possible form of $\Gamma_{div}^{(n)}$
reads
\begin{eqnarray}
\Gamma_{div}^{(n)} = \int d^3 x [ a_{(n)} \sqrt{- g} + b_{(n)} \sqrt{- g} R ].
\label{Div-EA}
\end{eqnarray}
These divergences can be absorbed by renormalizing the Newton's constant
and adding the cosmological constant's counter-term. Of course, after
renormalization, we take the limit $m \rightarrow 0$ and recover the
new massive gravity theory up to irrelevant free terms.

In this way, we have completed the proof of renormalizability of
the new massive gravity theory in three dimensions. One important remark
is that the higher derivative curvature-squared terms receive no quantum 
corrections since they have dimension of $(mass)^4$, thereby making
it possible for the new massive gravity to make sense of even after 
renormalization.

\section{Discussions}

In this article, we have presented a proof of renormalizability of a recently 
proposed new massive gravity theory in three dimensions \cite{Bergshoeff}. 
Our proof relies on the existence of the BRST symmetry and the BRST-invariant
infrared regulator.

Since it has been already shown that the new massive gravity theory is an interactive and
unitary theory for the massive gravitons of helicities $\pm 2$, our proof
insists that this theory is in addition renormalizable in the perturbation
theory. Thus, we have obtained a nontrivial toy model of perturbative quantum gravity. 

A peculiar feature of our proof is that we have adopted the Pauli-Fierz
mass term as the (infrared) regulator. Then, a natural question arises
whether or not this method could be applied to the proof of renormalizability of
the topologically massive gravity. We wish to clarify this problem in a future
publication.

\begin{flushleft}
{\bf Acknowledgement}
\end{flushleft}

We would like to thank M. Maeno for valuable discussions. 

\vs 1   

\end{document}